\documentclass[sigconf]{acmart}

\usepackage{booktabs} 
\usepackage{multirow}
\usepackage{subfigure}
\usepackage{verbatim}
\usepackage{CJKutf8}
\usepackage{flushend}

\usepackage{makecell}
\usepackage[utf8]{inputenc}
\setcopyright{rightsretained}

\usepackage{amsmath}

\usepackage{amssymb}
\usepackage{algorithm}
\usepackage{algorithmic}

\acmDOI{10.475/123_4}

\acmISBN{123-4567-24-567/08/06}

\acmConference[KDD 2021]{ACM SIGKDD Conference on Knowledge Discovery and Data Mining}{August 2021}{Singapore}
\acmYear{2021}
\copyrightyear{2021}
\acmArticle{4}
\acmPrice{15.00}

\editor{Jennifer B. Sartor}
\editor{Theo D'Hondt}
\editor{Wolfgang De Meuter}

\begin{document}

\title{FedGNN: Federated Graph Neural Network for   Privacy-Preserving Recommendation}
\fancyhead{}

\author{Chuhan Wu$^1$, Fangzhao Wu$^2$, Yang Cao$^3$, Yongfeng Huang$^1$, Xing Xie$^2$}

\affiliation{%
  \institution{$^1$Tsinghua University, Beijing 100084, China \\ $^2$Microsoft Research Asia, Beijing 100080, China,
  $^3$Kyoto University, Kyoto 615-8558, Japan
  }
} 
\email{{wuchuhan15, wufangzhao}@gmail.com, yang@i.kyoto-u.ac.jp, yfhuang@.tsinghua.edu.cn, xing.xie@microsoft.com}






\begin{abstract}
Graph neural network (GNN) is widely used for recommendation to model high-order interactions between users and items.
Existing GNN-based recommendation methods rely on centralized storage of user-item graphs and centralized model learning.
However, user data is privacy-sensitive, and the centralized storage of user-item graphs may arouse privacy concerns and risk.
In this paper, we propose a federated framework for privacy-preserving GNN-based recommendation, which can collectively train GNN models from decentralized user data and meanwhile exploit high-order user-item interaction information with privacy well protected.
In our method, we locally train GNN model in each user client based on the user-item graph inferred from the local user-item interaction data.
Each client uploads the local gradients of GNN to a server for aggregation, which are further sent to user clients for updating local GNN models.
Since local gradients may contain private information, we apply local differential privacy techniques to the local gradients to protect user privacy.
In addition, in order to protect the items that users have interactions with, we propose to incorporate randomly sampled items as pseudo interacted items for anonymity. 
To incorporate high-order user-item interactions, we propose a user-item graph expansion method that can find neighboring users with co-interacted items and exchange their embeddings for expanding the local user-item graphs in a privacy-preserving way.
Extensive experiments on six benchmark datasets validate that our approach can achieve competitive results with existing centralized GNN-based recommendation methods and meanwhile effectively protect user privacy.
\end{abstract}

%
%


\maketitle

\section{Introduction}

\begin{figure}[!t]
    \centering
    \subfigure[Centralized learning.]{  \label{fig:exp1}
    \includegraphics[width=0.4\linewidth]{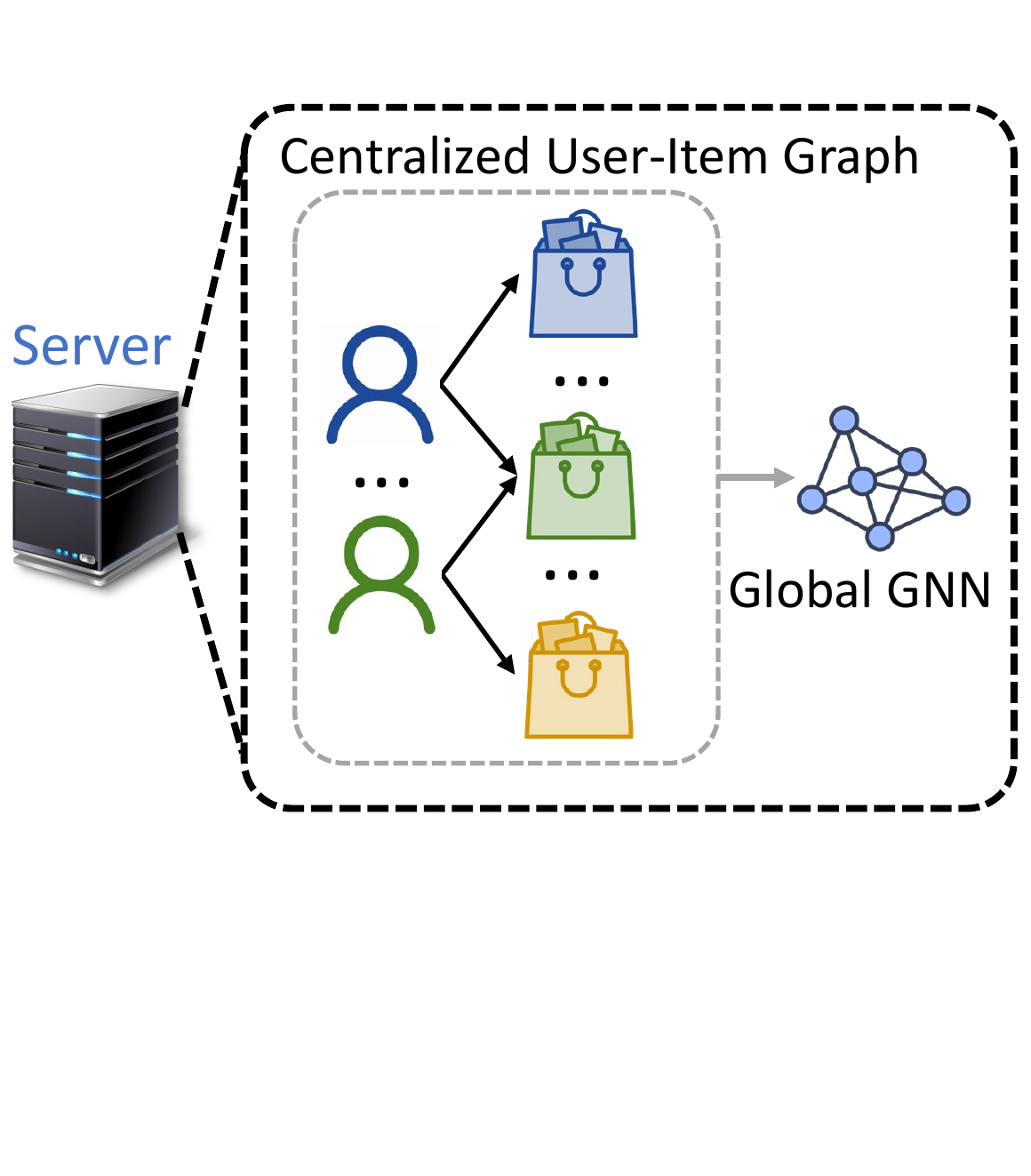} \vspace{0.5in}
    }
    \subfigure[Decentralized learning.]{  \label{fig:exp2}
    \includegraphics[width=0.53\linewidth]{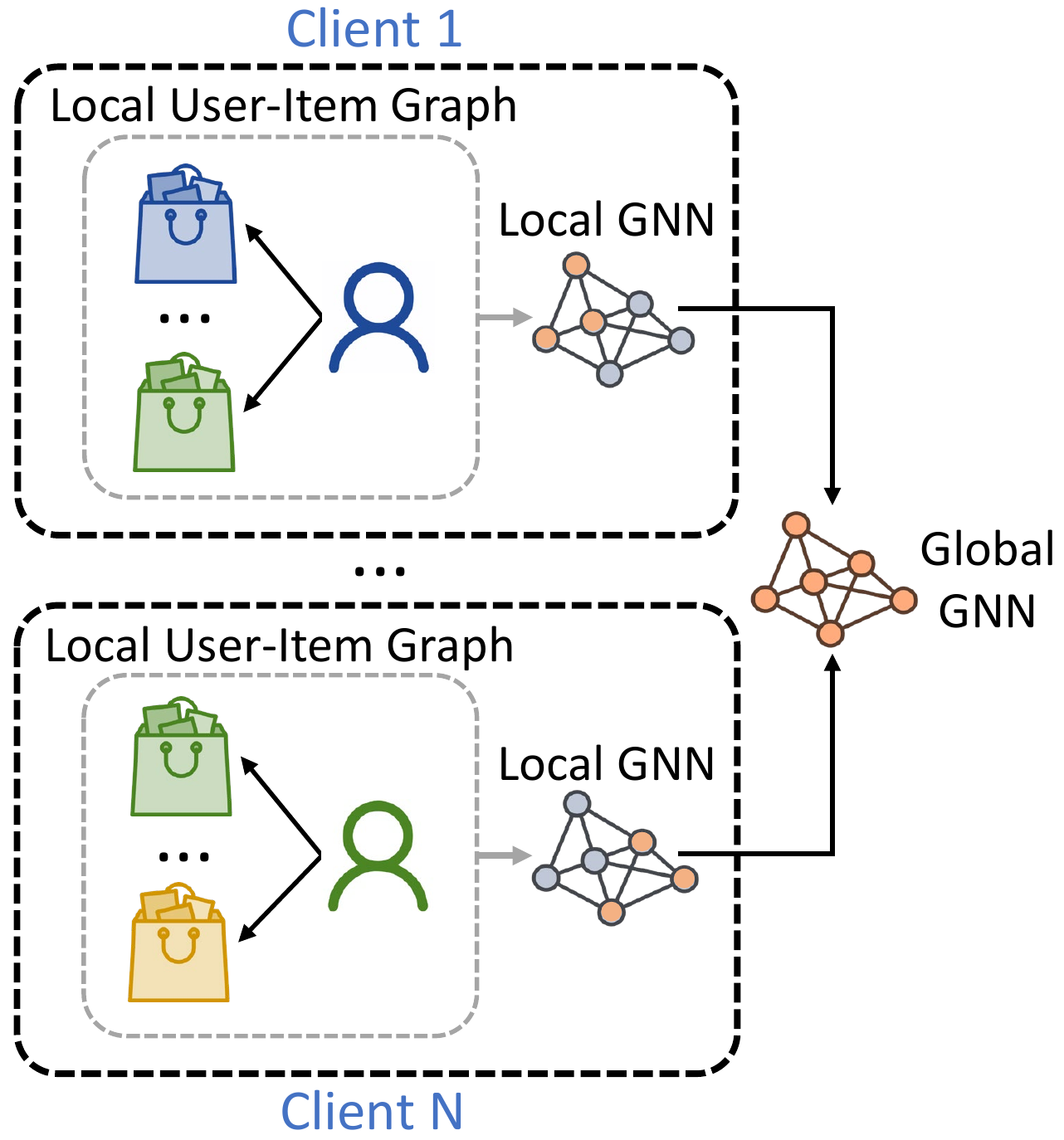} 
    }
\caption{Comparisons between centralized and decentralized training of GNN based recommendation models.}
    \label{fig:exp}
\end{figure}

Graph neural network (GNN) is widely used by many personalized recommendation methods in recent years~\cite{ying2018graph,wang2019neural,jin2020multi}, since it can capture high-order interactions between users and items on the user-item graph to enhance the user and item representations~\cite{berg2017graph,zhou2018graph,zhang2019star}.
For example, \citet{berg2017graph} proposed to use graph convolutional autoencoders to learn user and item representations from the user-item bipartite graph.
\citet{wang2019neural} proposed to use a three-hop graph attention network to capture the high-order interactions between users and items.
These existing GNN-based recommendation methods usually necessitate centralized storage of the entire user-item graph to learn GNN models and the representations of users and items, which means that the user-item interaction data needs to be centrally stored, as shown in Fig.~\ref{fig:exp1}.
However, user-item interaction data is highly privacy-sensitive, and its centralized storage can lead to the privacy concerns of users and the risk of data leakage~\cite{shin2018privacy}.
Moreover, under the pressure of strict data protection regulations such as GDPR\footnote{https://gdpr-info.eu}, online platforms may not be able to centrally store user-item interaction data to learn GNN models for recommendation in the future.

An intuitive way to tackle the privacy issue of user-item interaction data is to locally store the raw data on user devices and learn local GNN models based on it, as shown in Fig.~\ref{fig:exp2}.
However, in this scenario, it is very difficult to train an accurate GNN model for recommendation due to following reasons.
First, for most users the volume of interaction data on their devices is too small to locally train accurate GNN models.
Thus, a unified framework that coordinates a large number of user clients to collectively learn an accurate global GNN model from decentralized user data is required.
Second, the local GNN model trained on local user data may convey private information, and it is challenging to protect user privacy when synthesizing the global GNN model from the local ones.
Third, the local user data only contains first-order user-item interactions, and users' interaction items  cannot be directly exchanged due to privacy restrictions.
Thus, it is very challenging to exploit the high-order user-item interactions without privacy leakage.

In this paper, we propose a federated framework named \textit{FedGNN} for privacy-preserving GNN-based recommendation, which can effectively exploit high-order user-item interaction information by collectively
training GNN models for recommendation in a privacy-preserving way.
Since user interaction data is highly decentralized, there is no global user-item graph. 
Thus, in our method each user device locally learns a GNN model and the embeddings of user and items based on the user-item graph inferred from the local user-item interaction data on this device.
The user devices compute the gradients of models and user/item embeddings and upload them to a central server,  which aggregates the gradients from a number of users and distributes them to user devices for local updates.
However, both the items with non-zero gradients and the GNN model gradients contain private information.
Thus, we propose a privacy-preserving model update method to protect user-item interaction data without locally memorizing the full item set during model training.
More specifically, we apply local differential privacy  (LDP) techniques to the local gradients computed by user clients to  protect user privacy.
In addition, in order to protect the real items that user interacted with when uploading the gradients of item embeddings, we generate random embedding gradients of a certain number of randomly sampled pseudo interacted items.
Besides, to exploit high-order information of the user-item graph without leaking user privacy, we propose a privacy-preserving user-item graph expansion method that aims to find the neighbors of users with co-interacted items and exchange their embeddings to expand their local user-item graph.
In this way, high-order information of the user-item graph can be exploited by the GNN model to enhance user and item representations, and the private user-item interaction data do not leak.
We conduct massive experiments on six widely used benchmark datasets for recommendation, and the results show that our approach can achieve competitive results with existing centralized GNN-based recommendation methods and meanwhile effectively protect user privacy.

The major contributions of this paper are summarized as follows:
\begin{itemize}
    \item We propose a novel federated framework for privacy-preserving GNN-based recommendation that can exploit highly decentralized user data to collectively train GNN models.
 \item We propose to protect model gradients in model training with local differential privacy and propose a pseudo interacted item sampling technique to protect the items that users have interactions with.
  \item We propose a privacy-preserving user-item graph expansion method to exploit high-order user-item interactions from decentralized user data.

  \item Extensive experiments and analysis on six benchmark datasets show that our approach can achieve competitive results with existing centralized GNN-based recommendation methods and meanwhile protect user privacy.
\end{itemize}

\section{Related Work}

\subsection{GNN for Recommendation}
Graph neural networks are preferred by many recommendation methods to model high-order relations between users and items~\cite{berg2017graph,ying2018graph,wang2019neural,wang2019kgat,fan2019graph,wu2019session,zhang2019star,wu2019reviews,wang2020global,hu2020graph,tao2020mgat,ge2020graph}.
For example, \citeauthor{berg2017graph}~\shortcite{berg2017graph} proposed a graph convolutional matrix completion (GC-MC) approach.
GC-MC uses a graph convolutional encoder to learn user and item representations from the user-item bipartite graph, and then predicts unknown ratings via a bilinear decoder.
\citeauthor{ying2018graph}~\shortcite{ying2018graph}
proposed a graph convolutional neural network based method for recommendation named PinSage.
It learns item representations from an item-item graph via 2-hop graph convolutions, and uses these representations in downstream recommendation tasks.
\citeauthor{wang2019neural}~\shortcite{wang2019neural} proposed a neural graph collaborative filtering (NGCF) approach that uses 3-hop graph neural networks to learn user and item embeddings from the user-item bipartite graph.
Besides the user-item graph, several GNN-based recommendation methods also incorporate other kinds of graphs into recommendation, such as user-item-entity graph~\cite{wang2019knowledge} and user-user-item graph~\cite{fan2019graph}.
For example, \citeauthor{wang2019kgat}~\shortcite{wang2019kgat} proposed a knowledge-graph enhanced recommendation approach based on Knowledge Graph Attention Network (KGAT).
They use a 3-hop graph-attention network to learn user, item and entity representations from a heterogeneous graph, which is formed by linking entities in knowledge graphs with items in the user-item graph.
\citeauthor{fan2019graph}~\shortcite{fan2019graph} proposed a social recommendation approach named GraphRec.
They use graph attention networks to learn user and item embeddings from the user-item bipartite graph and the user-user social graph.
However, these methods need centralized storage of users' interactions with items to form the entire user-item graph, which may arouse users' privacy concerns and the risk of private data leakage. 
Different from them, in our \textit{FedGNN} method the raw user data never leaves the local user devices.
In addition, \textit{FedGNN} leverages a privacy-preserving model update method to protect private gradients and a privacy-preserving user-item graph expansion method to incorporate high-order user-item interactions.
Thus, \textit{FedGNN} can employ GNN models for grasping high-order information in a privacy-preserving way.

\begin{table*}[!t]
\centering

\caption{Comparison of different methods in terms of high-order user-item interaction modeling and privacy protection.  ``Cen.'' and ``Local'' represent centralized and decentralized data storage, respectively.}\label{table.result}
 
\resizebox{1.0\linewidth}{!}{
\begin{tabular}{lcccccccccc}
\Xhline{1.5pt}
              & PMF & SVD++ & GRALS & sRGCNN & GC-MC & PinSage & NGCF & FCF & FedMF & FedGNN \\ \hline
High-order user-item interaction   &   $\times$    &     \checkmark    &     \checkmark     & \checkmark     &      \checkmark    &       \checkmark   &      \checkmark      & $\times$    &   $\times$    &   \checkmark     \\
Rating protection   &   $\times$    & $\times$    &    $\times$     &    $\times$     &      $\times$    &      $\times$   &     $\times$      & \checkmark    &   \checkmark    &   \checkmark     \\
Interaction item protection &  $\times$    &  $\times$    &  $\times$     &    $\times$    &  $\times$     &  $\times$      &   $\times$       &  $\times$   &  $\times$     &    \checkmark     \\ 
User data storage &  Cen.    &  Cen.    &   Cen.     &    Cen.    &  Cen.    &  Cen.     &   Cen.       &  Local   &  Local     &  Local    \\ \Xhline{1.5pt}
\end{tabular}
}
\end{table*}

\subsection{Federated Learning}

Federated learning is a machine learning technique to collectively learn intelligent models based on decentralized user data in a privacy-preserving manner~\cite{konevcny2016federated,mcmahan2017communication}.
Different from existing machine learning methods based on centralized storage of user data, in federated learning the user data is kept locally on user devices~\cite{yang2019federated}.
Each device maintains a local model and computes local model updates based on the user data stored on this device.
The local model updates from a number of users are uploaded to a central server that coordinates the model training process.
These updates are aggregated into a unified one to update the global model maintained by this server.
The updated model is further distributed to all user devices to  update the local models.
This process is iteratively executed until the model converges.
Since the model updates usually contain much less private information and the raw user data never leaves the devices, the risk of privacy leakage can be effectively reduced~\cite{hard2018federated}.

\begin{figure*}[!t]
    \centering
    \includegraphics[width=1.0\linewidth]{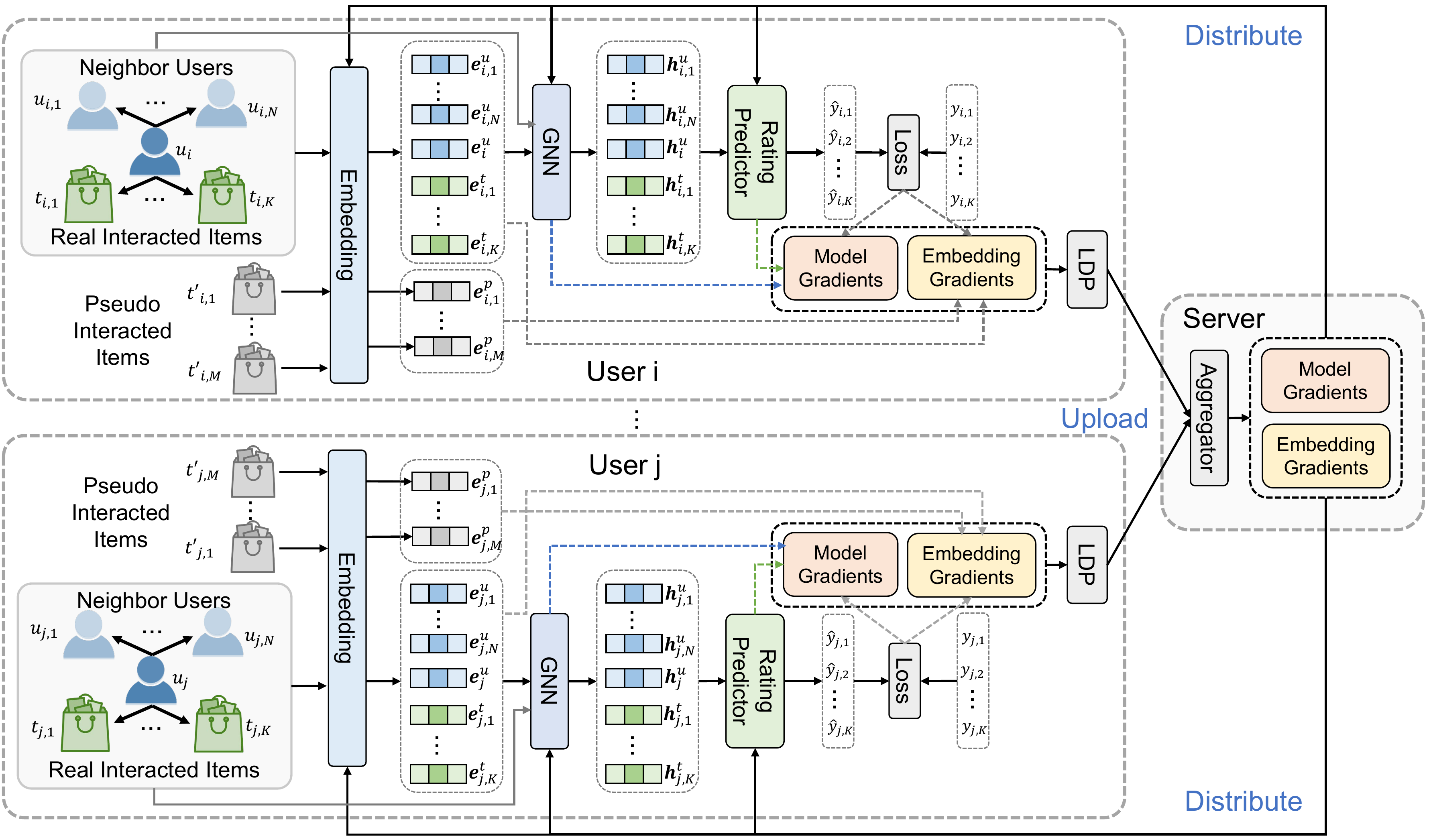} 
\caption{The framework of  our \textit{FedGNN} approach.}
 
    \label{fig:model3}
\end{figure*}

The framework of federated learning has been applied to personalized recommendation~\cite{ammad2019federated,chai2019secure,qi2020privacy,hegedHus2019decentralized,flanagan2020federated}.
For example, \citeauthor{ammad2019federated}~\shortcite{ammad2019federated} proposed a federated collaborative filtering (FCF) approach.
In FCF, each user device locally computes the gradients of the user and item embeddings based on the personal ratings stored on this device. 
The user embeddings are locally updated, and the gradients of item embeddings are uploaded to a central server.
The server aggregates the item gradients from massive clients to update the global item embeddings it maintains.
The updated item embeddings are further distributed to user clients for local embedding updates.
However, in this method the gradients of item embedding may leak some information on the private ratings~\cite{deepleakge}.
To solve this problem, \citeauthor{chai2019secure}~\shortcite{chai2019secure} proposed a federated matrix factorization  (FedMF) method, where the item embeddings are protected by homomorphic encryption techniques.
However, these methods do not consider the high-order interactions between users and items, which may not be optimal in learning accurate user and item representations.
In addition, they mainly focus on protecting the private  ratings given by users and cannot protect the raw user-item interaction data unless they locally maintain the full item set on each device, which is impractical due to the heavy storage and communication costs.
Different from these methods, our approach can capture high-order interactions between users and items by our proposed privacy-preserving user-item graph expansion method.
In addition, our method can protect the raw user-item interaction data during the model training process in an effective and efficient way.
To better demonstrate the advantage of our approach, we summarize the comparison between \textit{FedGNN} and existing methods on exploiting high-order user-item interactions and privacy protection in Table~\ref{table.result}.

\section{Methodology}
In this section, we first present the problem definitions in our federated GNN-based recommendation framework (FedGNN), then introduce the details of our \textit{FedGNN} approach for privacy-preserving recommendation, and finally provide some discussions and analysis on privacy protection.

\subsection{Problem Formulation}

Denote $\mathcal{U}=\{u_1, u_2, ..., u_P\}$ and $\mathcal{T} =\{t_1, t_2, ..., t_Q\}$ as the sets of users and items respectively, where $P$ is the number of users and $Q$ is the number of items. 
Denote the rating matrix between users and items as $\mathbf{Y}\in\mathbb{R}^{P \times Q}$, which is used to form a bipartite user-item graph $\mathcal{G}$ based on the observed ratings $\mathbf{Y}_o$. 
We assume that the user $u_i$ has interactions with $K$ items, which are denoted by $[t_{i,1}, t_{i,2}, ..., t_{i,K}]$.
These items and the  user $u_i$ can form a first-order local user-item subgraph $\mathcal{G}_i$.
The ratings that given to these items by user $u_i$ are denoted by $[y_{i,1}, y_{i,2}, ..., y_{i,K}]$.
To protect user privacy (both the private ratings and the items a user has interactions with), each user device locally keeps the interaction data of this user, and the raw data never leaves the user device.
We aim to predict the unobserved ratings ($y\in \mathbf{Y}\backslash \mathbf{Y}_o$) based on the interaction data $\mathcal{G}_i$ locally stored on user devices in a privacy-preserving way.
Note that there is no global user-item interaction graph  in our approach and local graphs are built and stored  in different device, which is very different from existing federated GNN methods~\cite{mei2019sgnn,jiang2020federated} that require the entire graph is built and stored together in at least one platform or device.

\subsection{FedGNN Framework}

Next, we introduce the framework of our \textit{FedGNN} method for privacy-preserving GNN-based recommendation.
It can leverage the highly decentralized user interaction data to learn GNN models for recommendation by exploiting the high-order user-item interactions in a privacy-preserving way.
The framework of \textit{FedGNN} is shown in Fig.~\ref{fig:model3}.
It mainly consists of a central server and a large number of user clients.
The user client keeps a local subgraph that consists of the user interaction histories with items and the neighbors of this user.
Each client learns the user/item embeddings and the GNN models from its local subgraph, and uploads the gradients to a central server.
The central server is responsible for coordinating these user clients in the model learning process by aggregating the gradients received from a number of user clients and delivering the aggregated gradients to them.
Next, we introduce how they work in detail.

The local subgraph on each user client is constructed from the user-item interaction data and the neighboring users that have co-interacted items with this user.
The node of this user is connected to the nodes of the items she interacted with, and the node of her neighboring users.
An embedding layer is first used to convert the user node $u_i$, the $K$ item nodes $[t_{i,1}, t_{i,2}, ..., t_{i,K}]$ and the $N$ neighboring user nodes $[u_{i,1}, u_{i,2}, ..., u_{i,N}]$ into their embeddings, which are  denoted as $\mathbf{e}^u_i$, $[\mathbf{e}^t_{i,1}, \mathbf{e}^t_{i,2}, ..., \mathbf{e}^t_{i,K}]$ and  $[\mathbf{e}^u_{i,1}, \mathbf{e}^u_{i,2}, ..., \mathbf{e}^u_{i,N}]$, respectively.
Since the user embeddings may not be accurate enough when the model is not well-tuned, we first exclude the neighboring user embeddings in the model learning for $T$ epochs, and then incorporate them into model learning when they have been tuned.
Note that the embeddings of the user $u_i$ and the item embeddings are can be locally tuned during model training, while the embeddings of neighboring users are fixed.\footnote{We find this method slightly outperforms using trainable neighboring user embeddings  (shown in experiments).  
Thus, we prefer fixed ones to reduce computational and communicational costs of model training.}

\begin{algorithm}[!t]
 \begin{algorithmic}[1]
 \STATE Each client constructs its local subgraph $\mathcal{G}_i$
   \STATE Initialize $\Theta_i$ on each user client using the same seed
   \STATE Iteration count $c\leftarrow0$
 \STATE Graph expansion switch $s\leftarrow True$
      \\[1.0ex]
    // Server\\[1.0ex]
    \REPEAT
    \STATE Select a subset $\mathcal{S}$ from the user set $\mathcal{U}$ randomly
    \STATE $\mathbf{g}=0$
    \FOR {each user client $u_i$ $\in$ $\mathcal{S}$} 
         \IF {$c|\mathcal{S}|<T\cdot P$}
         \STATE {$\mathbf{g} \leftarrow\mathbf{g}+\textbf{LocalGradCal}(\mathcal{G}_i, False)$}
         \ELSE {

         \STATE $\mathbf{g} \leftarrow\mathbf{g}+\textbf{LocalGradCal}(\mathcal{G}_i, True)$
                 } 
         \ENDIF
    \ENDFOR
    \IF {$c|\mathcal{S}|\geq T\cdot P$ and $s$}
             \STATE $\textbf{PrivacyPreservingGraphExpansion}()$
             \STATE $s\leftarrow False$
    \ENDIF
    \STATE $\mathbf{g} \leftarrow\mathbf{g}/|\mathcal{S}|$
    \STATE Distribute $\mathbf{g}$ to user clients for local update
    \UNTIL{model convergence}
\\[1.0ex]
    // User Client\\[1.0ex]
    
    \textbf{LocalGradCal}($i$, includeNeighbor):\\

    \STATE Select a mini-batch of data $\mathcal{N}$  from $\mathcal{G}_i$ 
    \IF {includeNeighbor}
    \STATE Use neighboring user embeddings
    \ENDIF
    \STATE Compute GNN model gradients $\mathbf{g}^m_i$ and embedding gradients $\mathbf{g}^e_i$ on  $\mathcal{N}$
    \STATE $\mathbf{g}_i\leftarrow (\mathbf{g}^m_i,\mathbf{g}^e_i)$ 
     \STATE \rm \textbf{return} $\mathbf{g}_i$
     \\[1.0ex]
     
\end{algorithmic}
    \caption{FedGNN}
\label{alg}
\end{algorithm}

Next, we apply a graph neural network to these embeddings to model the interactions between nodes on the local first-order sub-graph.
Various kinds of GNN network can be used in our framework, such as graph convolution network (GCN)~\cite{GCN}, gated graph neural network (GGNN)~\cite{ggnn} and graph attention network (GAT)~\cite{GAT}.
The GNN model outputs the hidden representations of the user and item nodes, which are denoted as $\mathbf{h}^u_i$,  $[\mathbf{h}^t_{i,1}, \mathbf{h}^t_{i,2}, ..., \mathbf{h}^t_{i,K}]$ and $[\mathbf{h}^u_{i,1}, \mathbf{h}^u_{i,2}, ..., \mathbf{h}^t_{i,N}]$, respectively.
Then, a rating predictor module is used to predict the  ratings given by the user $u_i$ to her interacted items (denoted by $[\hat{y}_{i,1}, \hat{y}_{i,2}, ..., \hat{y}_{i,K}]$) based on the embeddings of items and this user.
These predicted ratings are compared against the gold ratings locally stored on the user device to compute the loss function.
For the user $u_i$, the loss function $\mathcal{L}_i$ is computed as $\mathcal{L}_i=\frac{1}{K}\sum_{j=1}^K |\hat{y}_{i,j}-y_{i,j}|^2$.
We use the loss $\mathcal{L}_i$ to derive the gradients of the models and embeddings, which are denoted by $\mathbf{g}^m_i$ and $\mathbf{g}^e_i$, respectively.
These gradients will be further uploaded to the server for  aggregation.

The server aims to coordinate all user devices and compute the global gradients to update the model and embedding parameters in these devices.
In each round, the server awakes a certain number of user clients to compute gradients locally and send them to the server.  
After the server receiving the gradients from these users, the aggregator in this server will aggregate these local gradients into a unified one $\mathbf{g}$.\footnote{We use the FedAvg~\cite{mcmahan2017communication} algorithm to implement the aggregator.}
Then, the server sends the aggregated gradients to each client to conduct local parameter update.\footnote{Only sends the gradients of the model  and the corresponding user and item (including pseudo interacted ones) embeddings.}
Denote the parameter set in the $i$-th user device as $\Theta_i$.
It is updated by $\Theta_i=\Theta_i-\alpha \mathbf{g}$, where $\alpha$ is the learning rate.
This process will be iteratively executed until the model converges.
We summarize the framework of our \textit{FedGNN} method in Algorithm~\ref{alg}.
We will then introduce two modules for privacy protection in FedGNN, i.e., a privacy-preserving model update module (corresponding to Lines 9-11 in Algorithm~\ref{alg}) for protecting gradients in the model update and a privacy-preserving user-item graph expansion module (corresponding to Line 15 in Algorithm~\ref{alg}) to protect user privacy when modeling high-order user-item interactions.

\subsection{Privacy-Preserving  Model Update}

If we directly upload the GNN model and item embedding gradients, then there may be some privacy issues due to following reasons.
First, for embedding gradients, only the items that a user has interactions with have non-zero gradients to update their embeddings, and the server can directly recover the full user-item interaction history based on the non-zero item embedding gradients.
Second, besides the embedding gradients, the gradients of the GNN model and rating predictor may also leak private information of user histories and ratings~\cite{deepleakge}, because the GNN model gradients encode the preferences of users on items.
In existing methods such as FedMF~\cite{chai2019secure},  homomorphic encryption techniques are applied to gradients to protect private ratings.
However, in this method the user device needs to locally memorize the embedding table of the entire item set $\mathcal{T}$ and upload it in every iteration to achieve user interaction history protection, which is impractical due to the huge storage and communication costs during model training.

To tackle these challenges, we propose two strategies to protect user privacy in the model update process.
The first one is pseudo interacted item sampling.
Concretely, we sample $M$ items that the user has not interacted with\footnote{There are many sampling methods such as using the displayed items that have no interaction with a user. In our experiments we randomly sample items from the full item set for simulation. In addition, $M$ needs to be larger than $K$ to protect user privacy.},
and randomly generate their gradients $\mathbf{g}^p_i$ using a Gaussian distribution with the same mean and co-variance values with the real item embedding gradients.
The real embedding gradients $\mathbf{g}^e_i$ are combined with the pseudo item embedding gradients $\mathbf{g}^p_i$, and the unified gradient of the model and embeddings on the $i$-th user device (Line 26 in Algorithm~\ref{alg}) is modified as $\mathbf{g}_i=(\mathbf{g}^m_i, \mathbf{g}^e_i, \mathbf{g}^p_i)$.
The second one is local differential privacy.
Following~\cite{qi2020fedrec}, we clip the local gradients on user clients based on their L$\infty$-norm with a threshold $\delta$, and apply a local differential privacy (LDP)~\cite{choi2018guaranteeing} module with zero-mean Laplacian noise to the unified  gradients to achieve better user privacy protection, which are formulated as follows:
\begin{equation}\label{ldp}
    \mathbf{g}_i=clip(\mathbf{g}_i,\delta)+ Laplace(0,\lambda),
\end{equation}
where $\lambda$ is the strength of Laplacian noise.\footnote{The privacy budget $\epsilon$ can be bounded by $\frac{2\delta}{\lambda}$.}
The protected  gradients $\mathbf{g}_i$ are uploaded to the server for aggregation.

\begin{figure}[!t]
    \centering
    \includegraphics[width=1.0\linewidth]{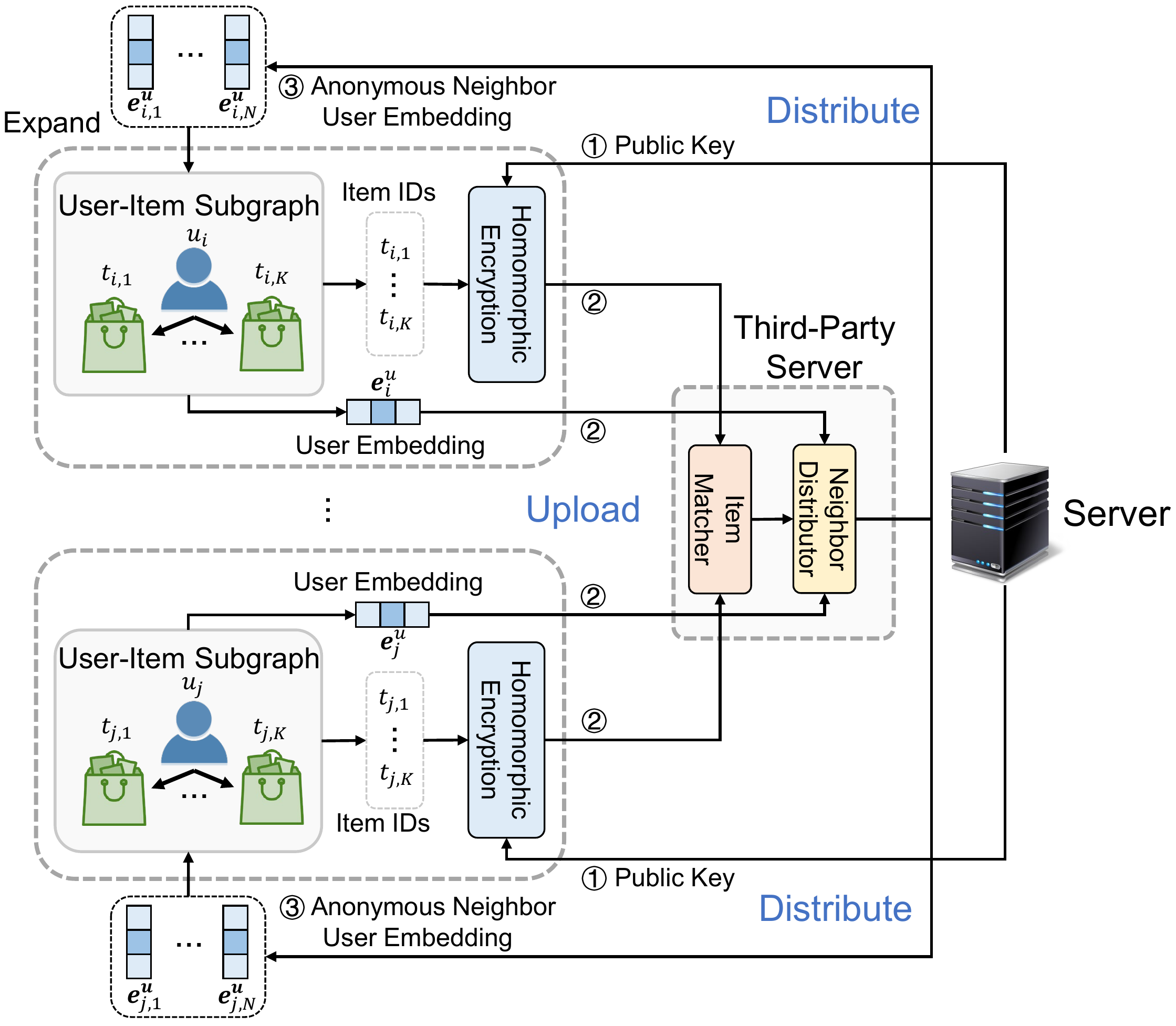}
\caption{The framework of the privacy-preserving user-item graph expansion  method.}

    \label{fig:model2}
\end{figure}

\subsection{Privacy-Preserving User-Item Graph Expansion}

Then, we introduce our privacy-preserving user-item graph expansion method that aims to find the neighbors of users and extend the local user-item graphs in a privacy-preserving way.
In existing GNN-based recommendation method based on centralized graph storage, high-order user-item interactions can be directly derived from the global user-item graph.
However, when user data is decentralized, it is a non-trivial task to incorporate high-order user-item interactions  without violating user privacy protection.
To solve this problem, we propose a privacy-preserving user-item graph expansion method that finds the anonymous neighbors of users to enhance user and item representation learning, where user privacy does not leak.
Its framework is shown in Fig.~\ref{fig:model2}.
We introduce its details as follows.

The central server that maintains the recommendation services first generates a public key, and then distributes it to all user clients for encryption.
After receiving the public key, each user device applies homomorphic encryption~\cite{chai2019secure} to the IDs of the items she interacted based on this key because the IDs of these items are privacy-sensitive.\footnote{We choose homomorphic encryption because the server cannot match the items hashed by many other salted encryption methods.}
The encrypted item IDs as well as the embedding of this user are uploaded to a third-party server (do not necessarily to be trusted).
This server finds the users who interacted with the same items via item matching, and then provides each user with the embeddings of her anonymous neighbors.
In this stage, the server for recommendation never receives the private information of users, and the third-party server cannot obtain any private information of users and items since it cannot decrypt the item IDs.\footnote{We assume that they do not collude with each other.}
We connect each user node with its neighboring user nodes.\footnote{The neighboring user nodes are not connected to the co-interacted items for better user privacy protection under Byzantine attack.}
In this way, the local user-item graph can be enriched by  the high-order user-item interactions without harming the protection of user privacy.
We summarize the process of our privacy-preserving   user-item graph expansion  method in Algorithm~\ref{alg2}.

\begin{algorithm}[!t]
 \begin{algorithmic}[1] 
\STATE  \textbf{PrivacyPreservingGraphExpansion}():
 \STATE Server sends a public key $p$ to user clients
 \STATE User clients encrypt item IDs with $p$
 \STATE User clients upload the user embedding and encrypted item IDs to a third-party server
 \STATE Third-party server distributes neighboring user embeddings to user clients
 \STATE User clients extend local graphs
\end{algorithmic}
    \caption{privacy-preserving user-item graph expansion }
\label{alg2}
\end{algorithm}

\subsection{Analysis on Privacy Protection}\label{sec:discussion}

The user privacy is protected by four aspects in our \textit{FedGNN} approach.
First, in \textit{FedGNN} the recommendation server never collects raw user-item interaction data, and only local computed gradients are uploaded to this server. 
Based on the data processing inequality~\cite{mcmahan2017communication}, we can infer that these gradients contain much less private information than the raw user interaction data.
Second, the third-party server also cannot infer private information from the encrypted item IDs since it cannot obtain the private key.
However, if the recommendation server colludes with the third-party server by exchanging the private key and item table, the user interaction history will not be protected.
Fortunately, the private ratings can still be protected by our privacy-preserving model update method.
Third, in \textit{FedGNN} we propose a pseudo interacted item sampling method to protect the real interacted items by sampling a number of items that are not interacted with a user.
Since gradients of both kinds of items have the same mean and co-variance values, it is difficult to discriminate the real interacted items from the pseudo ones if the number of pseudo interacted items is sufficiently larger than the number of real interacted items.
The average degree of privacy protection is proportional to $1+\frac{MP}{|Y_o|}$~\cite{sweeney2002k}.
Thus, the number of pseudo interacted items can be relatively larger to achieve better privacy protection as long as the computation resources of user devices permit.
Fourth, we apply the LDP technique to the gradients locally computed by the user device, making it more difficult to recover the raw user consumption history from these gradients.
It is shown in~\cite{qi2020fedrec} that the upper bound of the privacy budget $\epsilon$ is $\frac{2\delta}{\lambda}$, which means that we can achieve a smaller privacy budget $\epsilon$
 by using a smaller clipping threshold $\delta$ or a larger noise strength $\lambda$.\footnote{Smaller budget means better privacy protection.}
However, the accuracy of model gradients will also be affected if the privacy budget is too small. 
Thus, we need to properly choose both hyperparameters to balance  model performance and privacy protection.

\section{Experiments}
\subsection{Dataset and Experimental Settings}

In our experiments, following~\cite{berg2017graph} we use six widely used benchmark datasets for recommendation, including  MovieLens\footnote{https://grouplens.org/datasets/movielens/} (100K, 1M, and 10M), Flixster, Douban, and YahooMusic.
We use the preprocessed subsets of the Flixster, Douban, and YahooMusic datasets provided by~\cite{monti2017geometric}.\footnote{https://github.com/fmonti/mgcnn}
We denote the three versions of MovieLens as ML-100K, ML-1M and ML-10M respectively, and we denote YahooMusic as Yahoo.
The detailed statistics of these datasets are summarized in Table~\ref{dataset}.
\begin{table}[h]
\centering
\caption{Statistics of the datasets.}\label{dataset}
 
 \resizebox{1.0\linewidth}{!}{
\begin{tabular}{lllll}
\Xhline{1.5pt}
\multicolumn{1}{c}{Dataset} & \multicolumn{1}{c}{\textbf{\#Users}} & \multicolumn{1}{c}{\textbf{\#Items}} & \multicolumn{1}{c}{\textbf{\#Ratings}} & \multicolumn{1}{c}{\textbf{Rating Levels}} \\ \hline
Flixster                    & 3,000                                & 3,000                                & 26,173                                 & 0.5,1,...,5                                \\
Douban                      & 3,000                                & 3,000                                & 136,891                                & 1,2,...,5                                  \\
Yahoo                  & 3,000                                & 3,000                                & 5,335                                  & 1,2,...100                                 \\
ML-100K                     & 943                                  & 1,682                                & 100,000                                & 1,2,...,5                                  \\
ML-1M                       & 6,040                                & 3,706                                & 1,000,209                              & 1,2,...,5                                  \\
ML-10M                      & 69,878                               & 10,677                               & 10,000,054                             & 0.5,1,...,5                                \\ \Xhline{1.5pt}
\end{tabular}
} 

\end{table}

\begin{table*}[!t]
	\centering
	
\caption{Performance of different methods in terms of RMSE. Results of FedGNN and the best-performed baseline are in bold.}\label{table.result2}
\resizebox{0.6\linewidth}{!}{
\begin{tabular}{lcccccc}
\Xhline{1.5pt}
\multicolumn{1}{c}{Methods} & Flixster & Douban & Yahoo & ML-100K & ML-1M & ML-10M \\ \hline
PMF~\cite{mnih2008probabilistic}                                  & 1.375    & 0.886  & 26.6       & 0.965   & 0.883 & 0.856  \\
SVD++~\cite{koren2008factorization}                                & 1.155    & 0.869  & 24.4       & 0.952   & 0.860 & 0.834  \\
GRALS~\cite{rao2015collaborative}                                & 1.313    & 0.833  & 38.0       & 0.934   & 0.849 & 0.808  \\
sRGCNN~\cite{monti2017geometric}                               & 1.179    & 0.801  & 22.4       & 0.922   & 0.837 & 0.789  \\
GC-MC~\cite{berg2017graph}                                & \textbf{0.941}    & 0.734  & \textbf{20.5}       & \textbf{0.905}   & \textbf{0.832} & \textbf{0.777}  \\
PinSage~\cite{ying2018graph}                              & 0.945    & \textbf{0.732}  & 21.0       & 0.914   & 0.840 & 0.790  \\ 
NGCF~\cite{wang2019neural}                              & 0.954    & 0.742  & 20.9       & 0.916   & 0.833 & 0.779  \\ \hline
FCF~\cite{ammad2019federated}                                  & 1.064    & 0.823  & 22.9       & 0.957   & 0.874 & 0.847  \\
FedMF~\cite{chai2019secure}                                & 1.059    & 0.817  & 22.2       & 0.948   & 0.872 & 0.841  \\\hline
FedGNN                               & \textbf{0.989}    & \textbf{0.790}  & \textbf{21.1}       & \textbf{0.920}   & \textbf{0.848} & \textbf{0.803}  \\ \Xhline{1.5pt}
\end{tabular}
}
\end{table*}

In our experiments,  we use graph attention network (GAT)~\cite{GAT} as the GNN model, and use dot product to implement the rating predictor.
The user and item embeddings and their hidden representations learned by graph neural networks are 256-dim.
The epoch threshold $T$ is 2.
The gradient clipping threshold $\delta$ is set to 0.1, and the strength of Laplacian noise in the LDP module is set to 0.2 to achieve 1-differential privacy.
The number of pseudo interacted items is set to 1,000.
The number of users used in each round of model training is 128, and the total number of epoch is 3.
The ratio of dropout~\cite{srivastava2014dropout} is 0.2.
SGD is selected as the optimization algorithm, and its learning rate is 0.01.
The splits of datasets are the same as those used in~\cite{berg2017graph}, and these hyperparameters are selected according to the validation performance.
The metric used in our experiments is rooted mean square error (RMSE), and we report the average RMSE scores over the 10 repetitions.

\begin{figure*}[!t]
    \centering
    \includegraphics[width=0.99\linewidth]{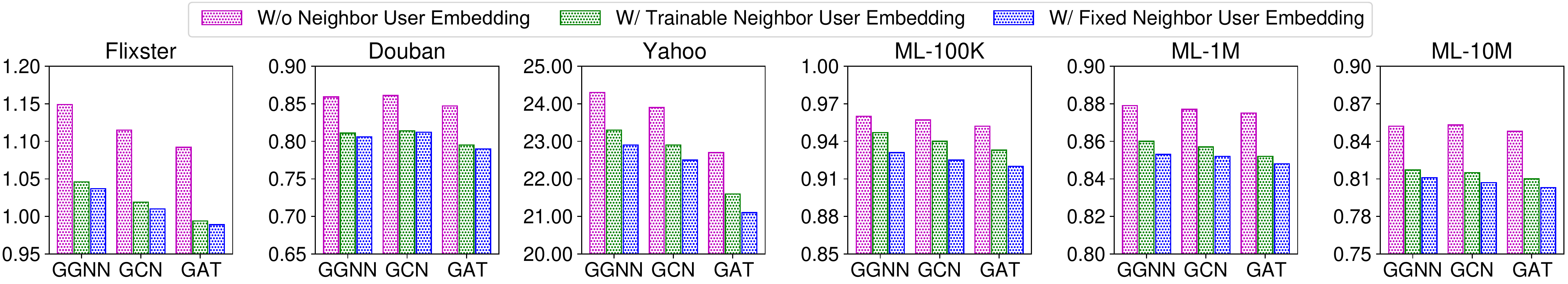} 
    \caption{Influence of second-order information and different GNN architectures.}
    \label{fig:effect}
\end{figure*}

\subsection{Performance Evaluation}
First, we compare the performance of our \textit{FedGNN} approach with several recommendation methods based on centralized storage of user data and several privacy-preserving ones based on federated learning, including:
\begin{itemize}
 \item  PMF~\cite{mnih2008probabilistic}, probability matrix factorization, which is a widely used recommendation method;
 \item SVD++~\cite{koren2008factorization}, another popular recommendation method based on a variant of singular value decomposition;
 \item  GRALS~\cite{rao2015collaborative}, a collaborative filtering approach with
graph information;
 \item  sRGCNN~\cite{monti2017geometric}, a matrix completion method with recurrent
multi-graph neural networks;
 \item  GC-MC~\cite{berg2017graph}, a matrix completion method based on graph convolutional autoencoders;
 \item  PinSage~\cite{ying2018graph}, a recommendation approach based on 2-hop GCN networks;
 \item  NGCF~\cite{wang2019neural}, a neural graph collaborative filter method;
 \item  FCF~\cite{ammad2019federated}, a privacy-preserving recommendation approach based on federated collaborative filtering;
 \item  FedMF~\cite{chai2019secure}, another privacy-preserving recommendation approach based on secure matrix factorization.
 
\end{itemize}

The recommendation performance of these methods is summarized in Table~\ref{table.result2}. 
We have several findings from Table~\ref{table.result2}.
First, we observe that the methods which incorporate high-order information of the user-item graph (e.g., \textit{GC-MC}, \textit{PinSage} and \textit{NGCF}) achieve better performance than those based on first-order information only (\textit{PMF}).
This is probably because modeling the high-order interactions between users and items can enhance user and item representation learning, and thereby improves the accuracy of recommendation.
Second, compared with the methods based centralized user-item interaction data storage like \textit{GC-MC} and \textit{NGCF}, our \textit{FedGNN} approach can achieve comparable or even better performance. 
It shows that our approach can protect user privacy and meanwhile achieve satisfactory recommendation performance.
Third, among the compared privacy-preserving recommendation methods, \textit{FedGNN} achieves the best performance.
This is because \textit{FedGNN} can incorporate high-order information of the user-item graphs, while \textit{FCF} and \textit{FedMF} cannot.
Besides, our approach can protect both ratings and user-item interaction histories, while \textit{FCF} and \textit{FedMF} can only protect ratings.

\begin{figure*}[!t]
    \centering
    \includegraphics[width=0.99\linewidth]{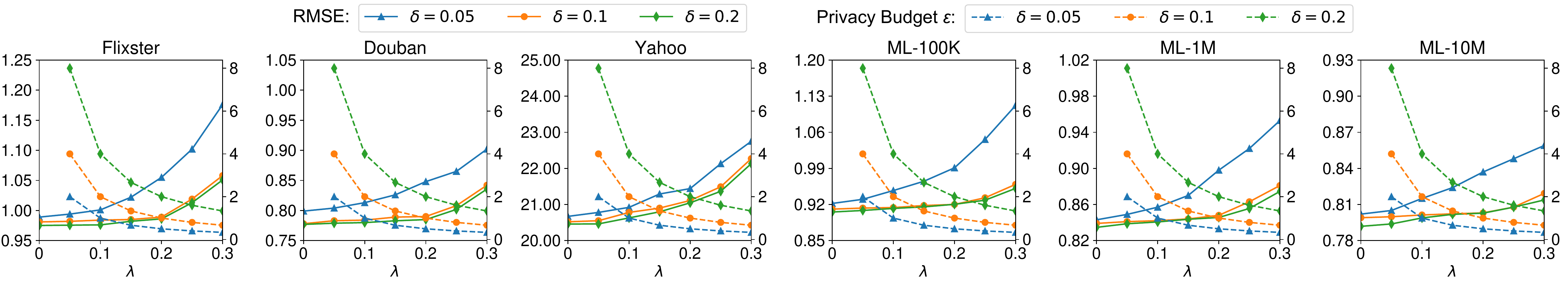} 
    \caption{The recommendation RMSE (left y-axis) and privacy budget $\epsilon$ (right y-axis) w.r.t. different clipping threshold $\delta$ and noise strength $\lambda$.} 
    \label{fig:lambda}
\end{figure*}

\begin{figure*}[!t]
    \centering
    \includegraphics[width=0.99\linewidth]{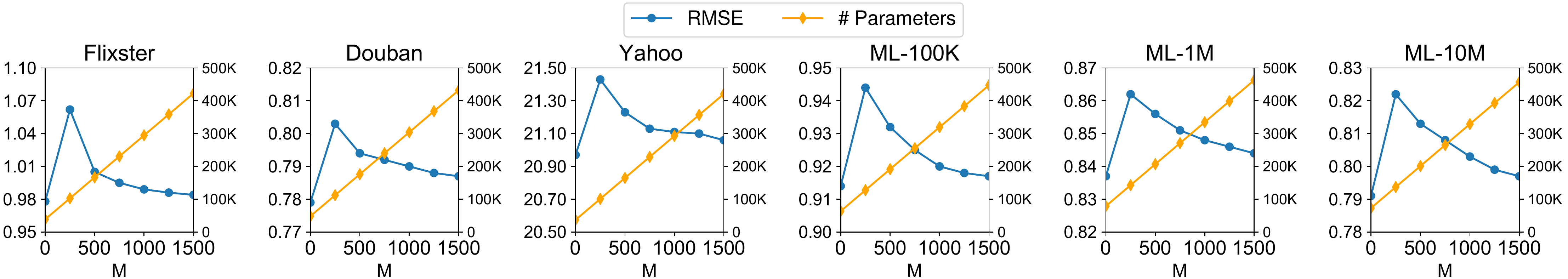} 
    \caption{The recommendation RMSE (left y-axis) and communication cost (right y-axis) w.r.t. different numbers of pseudo interacted items.} 
    \label{fig:M}
\end{figure*}

\subsection{Model Effectiveness}

Then, we validate the effectiveness of incorporating high-order information of the user-item graphs as well as the generality of our approach.
We compare the performance of \textit{FedGNN} and its  variants with fully trainable neighbor user embeddings or without high-order user-item interactions.
In addition, we also compare their results under different implementations of their GNN models (GGNN, GCN and GAT).
The results are shown in Fig.~\ref{fig:effect}, which reveals several findings.
First, compared with the baseline performance reported in Table~\ref{table.result2}, the performance of \textit{FedGNN} and its variants implemented with other different GNN models is satisfactory.
This result shows that our approach is compatible with different GNN architectures.
Second, \textit{FedGNN} slightly outperforms its variants based on GCN and GGNN.
This may be because the GAT network can more effectively model the importance of the interactions between nodes than GCN and GGNN, which is beneficial for user and item modeling. 
Third, the variants that can utilize the high-order information by using our \textit{FedGNN} framework perform better than those without high-order information.
It validates the effectiveness of our approach in incorporating high-order information of the user-item graph into recommendation.
Fourth, we find that using fixed neighbor user embeddings that are trained for a certain number of iterations is slightly better than using fully trainable ones that are updated in each iteration.
This may be because the neighboring user embeddings may not be accurate at the beginning of model training, which is not beneficial for learning precise user and item representations.

\subsection{Hyperparameter Analysis}

Finally, we explore the influence of three important hyperparameters, i.e., the gradient clip threshold $\delta$, the strength $\lambda$ of the Laplacian noise in the LDP module, and the number $M$ of the pseudo interacted items.
We first compare the performance of our \textit{FedGNN} approach by varying both hyperparameters, and the results are plotted in Fig.~\ref{fig:lambda}.\footnote{A larger $\lambda$ and smaller $\delta$ means smaller budget $\epsilon$, i.e., better privacy protection.} 
According to these results, we find that the difference between the model performance under  $\delta=0.1$ and $\delta=0.2$ is quite marginal.
However, if we clip the gradients with a smaller threshold such as 0.05, the prediction error will substantially increase.  
Thus, we prefer to set $\delta=0.1$ because we can achieve better privacy protection without much sacrifice of model performance.
In addition, the model performance declines with the growth of the noise strength $\lambda$, while the performance loss is not too heavy if $\lambda$ is not too large.
Thus, a moderate value of $\lambda$ such as 0.2 is preferable to achieve a good balance between privacy protection and recommendation accuracy.\footnote{We achieve 1-differential privacy under  $\delta=0.1$ and  $\lambda=0.2$.}

We also compare the performance and communication cost\footnote{We use the number of parameters to be exchanged in each iteration during model training to measure the communication cost.} of \textit{FedGNN} w.r.t. different $M$ in Fig.~\ref{fig:M}.
From Fig.~\ref{fig:M}, we observe that the performance is the best if $M$ is 0, but the user-item interaction histories cannot be protected.
According to the discussions in Section~\ref{sec:discussion}, if $M$ is too small the user privacy cannot be well-protected.
In addition, the performance also declines because the randomly generated gradients will influence the accuracy of item gradients.
By comparing the results on the three MovieLens datasets, we find that the rating matrix is sparser\footnote{The rating density of the ML-100K, 1M, and 10M datasets is 0.0630, 0.0447 and 0.0134, respectively.}, $M$ may need to be larger to keep good recommendation performance.
This ie because when $M$ is relatively large, the random gradients of pseudo interacted items will be better counteracted after aggregation and their influence will be mitigated.
However, as shown in Fig. \ref{fig:M}, the communication cost is also proportional to $M$ and it will be very heavy if $M$ is too large.
Therefore, we choose $M$ as 1,000 to achieve good privacy protection and recommendation performance under reasonable communication cost.

\section{Conclusion}

In this paper, we propose a federated framework for privacy-preserving GNN-based recommendation, which aims to  collectively train GNN models from decentralized user data by exploiting high-order user-item interactions in a privacy-preserving  manner. 
Concretely, we locally train GNN model in each user client based on the local user-item graph stored on this device.
Each client uploads the locally computed gradients to a server for aggregation, which are further sent to user clients for local updates.
In addition, to protect user-item interaction data during model training, we apply local differential privacy techniques to the local gradients to enhance user privacy protection.
Moreover, we sample pseudo interacted items to protect the embeddings of items that users have interactions with. 
Besides, to incorporate high-order user-item interaction information into model learning, we propose a privacy-preserving user-item graph expansion method that can find neighboring users with co-interacted items and exchange their embeddings for extending their local  graphs.
Massive experiments on six benchmark datasets show that our approach can achieve competitive performance with existing methods based on centralized storage of user-item interaction data and meanwhile effectively protect user privacy.

\bibliographystyle{ACM-Reference-Format}
\bibliography{main}

\end{document}